\documentclass[aps,preprint]{revtex4}%
\usepackage{amsfonts}
\usepackage{amsmath}
\usepackage{amssymb}
\usepackage{graphicx}
\usepackage{natbib}%
\providecommand{\U}[1]{\protect\rule{.1in}{.1in}}

\begin{document}
\preprint{ }
\title{Finite Temperature Quantum Effects on Confined Charges}
\author{Jeffrey Wrighton and James Dufty}
\affiliation{Department of Physics, University of Florida, Gainesville, FL 32611, USA}
\author{Sandipan Dutta}
\affiliation{Center for Soft and Living Matter, Department of Physics, Ulsan National Institute of Science and Technology, Ulsan 689-798, Republic of Korea}

\begin{abstract}
A quantum system of $N$ Coulomb charges confined within a harmonic trap is
considered over a wide range of densities and temperatures. A recently
described construction of an equivalent classical system is applied in order
to exploit the rather complete classical description of harmonic confinement via liquid state theory. Here, the effects of quantum mechanics on that representation
are described with attention focused on the origin and nature of shell
structure. The analysis extends from the classical strong Coulomb coupling
conditions of dusty plasmas to the opposite limit of low temperatures and
large densities characteristic of "warm, dense matter".

\end{abstract}
\date{\today}
\maketitle

\section{Introduction and Motivation}

\label{sec1}Coulomb correlations have been the focus of intense study for more
than fifty years. Weak coupling conditions, both classical and quantum, are
now well understood. The more interesting and difficult conditions of strong
Coulomb coupling are well understood only in the limiting cases of zero
temperature (electrons) and high temperatures (classical ions). Renewed
interest in the intermediate cross-over domain between quantum and classical
limits at arbitrary coupling has followed from new experimental studies of
\textquotedblleft warm, dense matter" \cite{WDM}, new theoretical approaches
\cite{PDW,DuftyDutta,DuttaDufty,DD13,Liu14}, and new path integral Monte Carlo
simulations \cite{Brown}. The objective here is to explore this domain of
finite temperatures for the case of charges in a harmonic trap under
conditions where confinement, strong coupling, and quantum effects can appear
together. Of particular interest is the role of these conditions in the
formation and characterization of shell structure.

The approach here is to exploit classical many-body methods that treat Coulomb
coupling effectively, such as classical density functional theory
\cite{Lutsko}, liquid state theory \cite{Hansen}, or molecular dynamics
simulation \cite{MD}. It is necessary first to embed relevant quantum effects
in a classical statistical mechanics. This has been shown to be an accurate
and practical idea recently by Perrot and Dharma-wardana \cite{PDW} using
liquid state theory, by introducing a pair potential modified to include
exchange and diffraction effects and an effective temperature to admit a
finite kinetic energy at zero temperature. This approach was formalized for a
more precise context by two of the current authors \cite{DuftyDutta}, and a
preliminary application to confined charges was described \cite{DuttaDufty}.
This effective liquid state approach has proved accurate for the
thermodynamics and structure of the uniform electron gas over a wide range of
densities and temperatures \cite{DD13,Liu14}. It is particularly useful for
the problem posed here since there is now a rather complete study of the
classical \textquotedblleft Coulomb balls" via liquid state theory and
classical Monte Carlo simulations \cite{Wrighton}. Once the effective quantum
potentials and thermodynamic parameters are specified, these same methods can be
applied directly to the questions of quantum effects on shell formation. That
is the objective of the work presented here.

At equilibrium the harmonically confined system is specified by the average
number of particles in the trap, $\overline{N}$, the temperature, $T$, and the
strength of the confining potential. The latter determines the volume of the
system (see below) so that ultimately the harmonic potential parameters can be
expressed in terms of the density and temperature. In the classical limit, all
density and temperature dependence of dimensionless quantities occurs only
through the classical Coulomb coupling constant, $\Gamma\equiv q^{2}%
/(r_{0}k_{B}T)$, where $q$ is the charge and $r_{0}$ is the Wigner-Seitz
length related to the average global density $\overline{n}$ by $r_{0}%
=(4\pi\overline{n}/3)^{-1/3}$. It is a measure of the Coulomb energy for a
pair of charges relative to the average kinetic energy per particle,
$q^{2}/k_{B}Tr=\Gamma/r^{\ast}$ where $r^{\ast}=r/r_{0}$. In the classical
case the primary results are that shell structure (peaks in the radial density
profile) appear only at sufficiently strong coupling ($\Gamma\gtrsim10$) and
sharpen as the coupling increases. The number of shells is determined entirely
by $\overline{N}$. A mean field description, without correlations, yields no
shell structure at any value of $\Gamma$. The equivalent classical system with
quantum effects has a different behavior. Initial study of a simple model
\cite{DuttaDufty} \ showed the emergence of a new origin for shell structure
even at weaker coupling due to exchange effects on the shape of the confining
potential. That simple model is reconsidered here in Section \ref{sec3}.
However, an improved model considered in Section \ref{sec4} shows that
mechanism to be significantly diminished \cite{WrightonDuftyDutta}.\ The
objective here is to explore the onset and competition for all of the
potential origins for shell structure - Coulomb correlations, diffraction,
exchange - as a function of the dimensionless density parameters $r_{s}%
=r_{0}/a_{b}$ (where $a_{b}$ is the Bohr radius in terms of the charge and
mass of the confined particles) and $t=k_{B}T/e_{F}$ (where $e_{F}$ is the
ideal gas Fermi energy per particle, again in terms of the confined particle's mass).

\begin{figure}[ptb]
\includegraphics[width=\columnwidth]{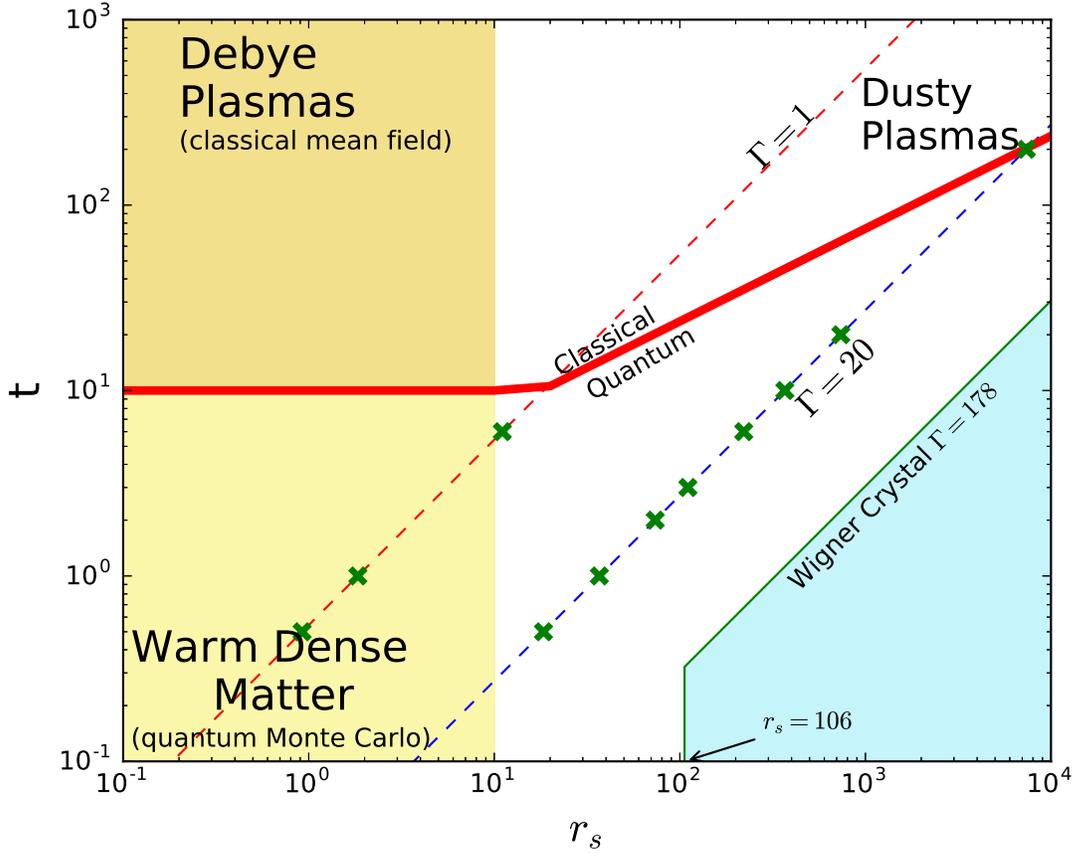}\caption{Values of $r_{s}$ and $t$
of interest. Note that the values correspond to a range of experimental
conditions from electrons to dusty plasmas. A definition of the crossover to
quantum effects from classical behavior is shown. Crosses indicate the
conditions studied in sections \ref{sec3} and \ref{sec4}.}%
\label{Fig.1}%
\end{figure}

To explore the full range of systems of interest requires a wide range of
values for $t$ and $r_{s}$. The upper limits are primarily imposed by the conditions
of strong coupling for classical shell structure, as occurs in dusty plasmas.
This is illustrated in Figure \ref{Fig.1}. For $r_{s}<10$ Coulomb effects are
weaker and the classical - quantum transition is dominated by $t$, for ideal
gas diffraction and exchange effects. Here the classical domain has been
defined as $t>10$. In contrast, for larger $r_{s}$ quantum effects on Coulomb
correlations dominate at higher $t$ and the coupling strength $\Gamma$ is
changed to an effective value $\Gamma_{e}\left(  t,r_{s}\right)  \leq\Gamma$
(see eq. $\left(  \ref{2.12}\right)  $ below). The classical limit in this
domain is defined to be $\Gamma_{e}/\Gamma>0.99.$ Typical experimentally
accessible values for electrons are $r_{s}<10$ over a wide range of
temperatures. This is the domain of zero temperature condensed matter physics,
warm dense matter, and Debye plasmas in the left side of Figure \ref{Fig.1}.
At the opposite extreme are the strongly coupled classical plasmas in the
upper right side of the figure. These large values of $t$ and $r_{s}$ can \ be
realized only for particles of large mass and charge, e.g. dusty plasmas
\cite{Bonitz10}. Intermediate domains are the primary interest here. The
constant $\Gamma$ lines are shown for $\Gamma=1$ and $20.$ The crosses on
these lines indicate values of $t,r_{s}$ for which calculations are reported
here. Since the parameter space is large only the case $\overline{N}=100$ is
considered. Also, only the fluid phase for unpolarized charges is considered;
for the crystal phase see reference \cite{Bonitz08}.

The next section defines the effective classical description for the density
profile in terms of the modified pair potential and confining potential - all
quantum effects occur through modifications of the underlying Coulomb and
harmonic forms, respectively. The approximate form for the pair potential is
described in Appendix \ref{apA}. As noted above it has been shown to give good
predictions for the pair correlation function of the uniform electron gas, in
comparison to quantum Monte Carlo simulation \cite{DD13}. The choice for the
modified confining potential is described in \ref{apB}, where the potential is
represented in terms of a "trial" quantum density imposing a known limit.
Density profiles calculated on the basis of chosen quantum input are given in
Sections \ref{sec3} and \ref{sec4} for values of $t$ and $r_{s}$ corresponding to the line $\Gamma=20$ in Figure \ref{Fig.1}. The purely
classical profile would be the same in all of these cases since it depends
only on $\Gamma$. Hence the observed differences are purely quantum effects.
Two choices for determination of the effect trap are explored here. The first
is that whose trial density is the limit of non-interacting Fermions in a
harmonic trap. At the highest values of $t$ and $r_{s}$ the classical limit
is valid and at $\Gamma=20$ Coulomb correlations are strong enough for shell
structure, well-known for dusty plasmas. At the smallest values of
$t$ and $r_{s}$ a different shell structure emerges from extreme distortion
of the non-interacting trial density due to exchange effects. The analysis for
a second choice of the effective trap is repeated in Section \ref{sec4} with
an improved trial density to include the effects of Coulomb interactions. With
this quantum input, the new shell structure at small $t$ and $r_{s}$ no
longer dominates and the quantum differences from the classical form are
quantitative rather than qualitative. This sensitivity of the classical theory
to the modifications of the confining potential, the need for guidance from
simulation, and the outlook for future applications in materials sciences are
discussed in the last section.

\section{ Density Profile - Classical Map of the Quantum System}

\label{sec2}The Hamiltonian for $N$ particles with charge $q$ in a harmonic
trap is
\begin{equation}
H-\mu N=\sum_{i=1}^{N}\frac{p_{i}^{2}}{2m}+\frac{1}{2}\sum_{i\neq j}^{N}%
\frac{q^{2}}{\left\vert \mathbf{r}_{i}-\mathbf{r}_{j}\right\vert }-\int
d\mathbf{r}\mu(\mathbf{r})\widehat{n}(\mathbf{r}), \label{2.1}%
\end{equation}
with the local chemical potential given explicitly as%
\begin{equation}
\mu(\mathbf{r})=\mu-\frac{1}{2}m\omega^{2}r^{2}, \label{2.2}%
\end{equation}
and the operator $\widehat{n}(\mathbf{r})$ representing the microscopic
density is%
\begin{equation}
\widehat{n}(\mathbf{r})=\sum_{i=1}^{N}\delta\left(  \mathbf{r}-\mathbf{q}%
_{i}\right)  . \label{2.2a}%
\end{equation}
The constant $\mu$ determines the average number of charges $\overline{N}$ at
equilibrium in the grand canonical ensemble. As a consequence of the harmonic
potential the equilibrium average density profile for the charges is
non-uniform%
\begin{equation}
n(\mathbf{r},\beta\mid\mu)=\Omega^{-1}\sum_{N=1}^{\infty}N\int d\mathbf{r}%
_{2}..d\mathbf{r}_{N}\left\langle \mathbf{r}..\mathbf{r}_{N}\right\vert
e^{-\beta\left(  H-\mu N\right)  }\left\vert \mathbf{r}..\mathbf{r}%
_{N}\right\rangle , \label{2.3}%
\end{equation}
where $\left\langle \mathbf{r}_{1}..\mathbf{r}_{N}\right\vert X\left\vert
\mathbf{r}_{1}..\mathbf{r}_{N}\right\rangle $ is the $N$ particle diagonal,
properly symmetrized (Fermions or Bosons) matrix element in coordinate
representation, and $\Omega$ is the grand potential%
\begin{equation}
\Omega(\beta\mid\mu)=\sum_{N=1}^{\infty}\int d\mathbf{r}_{1}..d\mathbf{r}%
_{N}\left\langle \mathbf{r}_{1}..\mathbf{r}_{N}\right\vert e^{-\beta\left(
H-\mu N\right)  }\left\vert \mathbf{r}_{1}..\mathbf{r}_{N}\right\rangle ,
\label{2.4}%
\end{equation}
The notation $f(a,b\mid c)$ indicates a function of the parameters $a,b$ and a
functional of $c\left(  \mathbf{r}\right)  $. The density profile in the
classical limit has been studied in detail, via simulation and theory
\cite{Wrighton}. In that case the dimensionless form depends only on
$\overline{N}$ and the Coulomb coupling constant $\Gamma=\beta q^{2}/r_{0}$.
For sufficiently large Coulomb coupling, $\Gamma,$ the formation of shell
structure is observed in $n(\mathbf{r})$. The objective here is to exploit
this classical description to explore the effects of quantum diffraction and
exchange via a proposed equivalent classical system
\cite{DuftyDutta,DuttaDufty}. The equivalent classical system has an effective
local chemical potential, $\mu_{c}(\mathbf{r})$, an effective pair potential,
$\phi_{c}(\left\vert \mathbf{r}_{i}\mathbf{-r}_{j}\right\vert )$, and an
effective inverse temperature, $\beta_{c}$. These must be given as functions
of $\mu(\mathbf{r})$, $\phi(\left\vert \mathbf{r}_{i}\mathbf{-r}%
_{j}\right\vert )$, and $\beta$ for the quantum system

The basis for the classical study used here is the hypernetted chain (HNC)
description for an inhomogeneous equilibrium system \cite{Attard89}, or Eq.
(37) of reference \cite{DD13}%

\begin{equation}
\ln\left(  n\left(  \mathbf{r},\beta_{c}\mid\mu_{c}\right)  \lambda_{c}%
^{3}\right)  =\beta_{c}\mu_{c}(\mathbf{r})+\int d\mathbf{r}^{\prime}%
c^{(2)}(\mathbf{r,r}^{\prime},\beta_{c}\mid\mu_{c})n\left(  \mathbf{r}%
^{\prime}\right)  , \label{2.5}%
\end{equation}
where $\lambda_{c}=\left(  2\pi\beta_{c}\hbar^{2}/m\right)  ^{1/2}$ is the
thermal de Broglie wavelength expressed in terms of the effective classical
temperature, and $c^{(2)}(\mathbf{r,r}^{\prime},\beta_{c}\mid\mu_{c})$ is the
direct correlation function defined by the Ornstein-Zernicke equation in terms
of the pair correlation function for the inhomogeneous system \cite{Attard89}.
Further details of the origins for this equation in classical density
functional theory are given in reference \cite{Wrighton}. The classical
studies made a further approximation to this expression, replacing the direct
correlation function for the inhomogeneous system by that for a corresponding
uniform one component plasma (OCP or jellium), $c^{(2)}(\mathbf{r,r}^{\prime
},\beta_{c}\mid\mu_{c})\rightarrow c(\left\vert \mathbf{r-r}^{\prime
}\right\vert ,\beta_{c},\mu_{c})$. The results based on this approximation are
found to be quite accurate except at very strong coupling. A partial
theoretical basis for this approximation has been given \cite{Wrighton12} and
it will be made here as well.

An equivalent Boltzmann form for the density is defined in terms of a
dimensionless potential $U(\mathbf{r})$ defined by
\begin{equation}
n\left(  \mathbf{r},\mu_{c},\beta_{c}\right)  =\overline{N}\frac
{e^{-U(\mathbf{r},\mu_{c},\beta_{c})}}{\int d\mathbf{r}^{\prime}%
e^{-U(\mathbf{r}^{\prime},\mu_{c},\beta_{c})}}, \label{2.6}%
\end{equation}
where (\ref{2.5}) gives
\begin{equation}
U(\mathbf{r,}\mu_{c},\beta_{c})=-\nu_{c}(\mathbf{r})-\frac{\overline{N}}{\int
d\mathbf{r}^{\prime}e^{-U(\mathbf{r}^{\prime},\mu_{c},\beta_{c})}}\int
d\mathbf{r}^{\prime}e^{-U(\mathbf{r}^{\prime},\mu_{c},\beta_{c})}c(\left\vert
\mathbf{r-r}^{\prime}\right\vert ,\mu_{c},\beta_{c}). \label{2.7}%
\end{equation}
The dimensionless activity, $\nu_{c}(\mathbf{r,}\mu_{c},\beta_{c})=\beta
_{c}\mu_{c}(\mathbf{r})$, has been introduced in (\ref{2.7}) and $c(r,\mu
_{c},\beta_{c})$ is now the direct correlation function for the uniform OCP.
For future reference, note that at fixed $\overline{N}$ the representation for
$n\left(  \mathbf{r}\right)  $ is invariant to a shift of $\nu_{c}%
(\mathbf{r})$ by a constant. In the following applications this flexibility
will be used to choose $U(\mathbf{0,}\mu_{c},\beta_{c})=0.$

Equations (\ref{2.6}) and (\ref{2.7}) are a classical representation for the
density profile (\ref{2.3}) for the underlying quantum system. The latter is
parameterized by the total average number of particles $\overline{N}$, the
inverse temperature $\beta$, and the chemical potential of the uniform system
$\mu$. In the following, a change of variables from $\beta,\mu$ to
$\beta,\overline{n}$ is considered, where $\overline{n}$ is the average density of the
representative uniform system. To introduce the density, it is necessary to
assign a volume for the system. This can be taken as the volume of a sphere
with radius $R_{0}$ corresponding to a particle at the greatest distance from
the center. At equilibrium the average fluid phase density is spherically
symmetric so that the total average force on that particle is%
\begin{equation}
\frac{\left(  \overline{N}-1\right)  q^{2}}{R_{0}^{2}}-m\omega^{2}%
R_{0}=0,\hspace{0.25in}\Rightarrow R_{0}^{3}=\left(  \overline{N}-1\right)
\frac{q^{2}}{m\omega^{2}}. \label{2.8}%
\end{equation}
This gives the average density to be
\begin{equation}
\overline{n}\equiv\frac{3\overline{N}}{4\pi R_{0}^{3}}=\frac{3m\omega^{2}%
}{4\pi q^{2}}\frac{\overline{N}}{\overline{N}-1}. \label{2.9}%
\end{equation}

As expected the density is determined by the trap parameter $m\omega^{2}%
/q^{2}$. A corresponding length scale $r_{0}$ is the average distance between
particles given by $4\pi r_{0}^{3}/3=1/\overline{n}$. The following
dimensionless measures of distance, temperature, and density will be used%
\begin{equation}
\mathbf{r}^{\ast}=\frac{r}{r_{0}},\hspace{0.2in}t=\frac{1}{\beta\epsilon_{F}%
},\hspace{0.2in}r_{s}=\frac{r_{0}}{a_{b}}. \label{2.10}%
\end{equation}
Here $\epsilon_{F}$ is the Fermi energy and $a_{b}$ is the Bohr radius, both
defined in terms of the mass and charge of the particles in the trap%
\begin{equation}
\epsilon_{F}=\frac{1}{2m}\hbar^{2}\left(  3\pi^{2}\overline{n}\right)
^{2/3}=\left(  \frac{m_{e}}{m}\right)  \epsilon_{eF},\hspace{0.2in}a_{b}%
=\frac{\hbar^{2}}{mq^{2}}=\left(  \frac{m_{e}e^{2}}{mq^{2}}\right)  a_{B},
\label{2.10a}%
\end{equation}%
\begin{equation}
\nu(\mathbf{r,}\mu_{e},\beta)=\beta\mu_{e}-\frac{1}{2}\Gamma(t,r_{s})r^{\ast2}
\label{2.10b}%
\end{equation}
In the last equalities of (\ref{2.10a}) $\epsilon_{eF}$ and $a_{B}$ are the
electron Fermi energy and the usual Bohr radius, respectively. The prefactor
$m_{e}e^{2}/mq^{2}$ shows how the very large values of $r_{s}$ in Figure
\ref{Fig.1} can be obtained for particles of large mass and large charge.

Finally, define the reduced potential $u(\mathbf{r}^{\ast},t,r_{s})$, direct
correlation function $\overline{c}(r^{\ast},t,r_{s})$, and local activity
$\overline{\nu}_{c}(\mathbf{r}^{\ast},t,r_{s})$ by%
\begin{equation}
U(\mathbf{r})=\Gamma_{e}(t,r_{s})u(\mathbf{r}^{\ast},t,r_{s}), \hspace{0.2in}
c(r,\mu_{c},\beta_{c})=\Gamma_{e}(t,r_{s})\overline{c}(r^{\ast}%
,t,r_{s}) , \hspace{0.2in}
\nu_{c}(\mathbf{r},\mu_{c},\beta_{c})=\Gamma
_{e}(t,r_{s})\overline{\nu}_{c}(\mathbf{r}^{\ast},t,r_{s}). \label{2.11}%
\end{equation}
An effective coupling constant $\Gamma_{e}(t,r_{s})$ has been extracted in
each case%
\begin{equation}
\Gamma_{e}\left(  t,r_{s}\right)  =\frac{2}{\beta\hbar\omega_{p}\coth\left(
\beta\hbar\omega_{p}/2\right)  }\Gamma,\hspace{0.2in}\Gamma\equiv\frac{\beta
q^{2}}{r_{0}}. \label{2.12}%
\end{equation}
Here $\omega_{p}=\sqrt{4\pi \overline{n}q^{2}/m}$ is the plasma frequency. The
dimensionless parameter is $\beta\hbar\omega_{p}=\left(  4/3\right)  \left(
2\sqrt{3}/\pi^{2}\right)  ^{1/3}\sqrt{r_{s}}/t$ $\simeq0.940\,52\sqrt{r_{s}%
}/t$ . At fixed $r_{s}$ and large $t$, $\Gamma_{e}\left(  t,r_{s}\right)
\rightarrow\Gamma\simeq0.543r_{s}/t$ which is the classical Coulomb coupling
constant. The motivation for introducing $\Gamma_{e}\left(  t,r_{s}\right)  $
is the fact that it represents the strength of the Coulomb tail for the
effective classical pair potential \cite{DuttaDufty}, as shown in Appendix
\hbox{\ref{apA} eq. (\ref{a.5})}. This means that the strength of the effective
classical repulsion of particles in the trap is $\Gamma_{e}\left(
t,r_{s}\right)  $ while the strength of the harmonic containment is
$\Gamma(t,r_{s})$ (see (\ref{2.10b})). Since $\Gamma_{e}\left(  t,r_{s}\right)
$ decreases with increasing quantum effects, stronger confinement relative to
the purely classical result is expected.

The dimensionless form for the density profile, from (\ref{2.6}) and
(\ref{2.7}) is now
\begin{equation}
n^{\ast}\left(  \mathbf{r}^{\ast},t,r_{s}\right)  =n\left(  \mathbf{r},\mu
_{c},\beta_{c}\right)  r_{0}^{3}=\overline{N}\frac{e^{-\Gamma_{e}%
(t,r_{s})u(\mathbf{r}^{\ast},t,r_{s})}}{\int d\mathbf{r}^{\ast\prime
}e^{-\Gamma_{e}(t,r_{s})u(\mathbf{r}^{\ast\prime},t,r_{s})}}, \label{2.13}%
\end{equation}%
\begin{equation}
u(\mathbf{r}^{\ast},t,r_{s})=-\overline{\nu}_{c}(\mathbf{r}^{\ast}%
,t,r_{s})-\frac{\overline{N}}{\int d\mathbf{r}^{\ast\prime\prime}%
e^{-\Gamma_{e}(t,r_{s})u(\mathbf{r}^{\ast\prime\prime},t,r_{s})}}\int
d\mathbf{r}^{\ast\prime}e^{-\Gamma_{e}(t,r_{s})u(\mathbf{r}^{\ast\prime
},t,r_{s})}\overline{c}(\left\vert \mathbf{r}^{\ast}\mathbf{-r}^{\ast\prime
}\right\vert ,t,r_{s}). \label{2.14}%
\end{equation}
Practical application requires specification of the direct correlation
function $\overline{c}(r^{\ast},t,r_{s})$ for jellium and the classical
activity $\overline{\nu}_{c}(r^{\ast},t,r_{s})$. The method for determining
these is such that they are explicit functions of the dimensionless variables
$t,r_{s}$ for the given quantum system, rather than of the associated
classical parameters $\mu_{c},\beta_{c}$. Hence, the potentially confusing
notation in (\ref{2.11}). The former is determined from an accurate equivalent
classical calculation described elsewhere \cite{DD13} and summarized in
Appendix \ref{apA}. The direct correlation function is a classical concept
whose quantum modifications here appear only through the effective pair
potential. That potential is obtained in Appendix \ref{apA} and has two main
changes from the underlying Coulomb potential due to quantum effects in the
classical representation. The first is a regularization of the Coulomb
singularity at the origin due to diffraction effects - the pair potential
remains finite at zero separation. The second main change is the strength of
the $1/r$ behavior at large distances, with the coupling constant $\Gamma$
being replaced by $\Gamma_{e}$ of (\ref{2.12}).

The activity $\overline{\nu}_{c}(\mathbf{r}^{\ast},t,r_{s})$ describes the
effective classical trap potential corresponding to the actual quantum
harmonic trap, and its approximate determination is described in Appendix
\ref{apB}. It is defined such that the density profile for a chosen quantum
system is recovered in an appropriate limit. In this way the exact\ quantum
effects of that limit are incorporated in the classical system and exploited
approximately away from that limit as well. The resulting form for (\ref{2.13})
and (\ref{2.14}) obtained in Appendix \ref{apB} is%
\begin{equation}
n^{\ast}\left(  \mathbf{r}^{\ast},t,r_{s}\right)  =N\frac{n_{T}^{\ast}\left(
\mathbf{r}^{\ast},t,r_{s}\right)  e^{\Gamma_{e}(t,r_{s})\Delta u(\mathbf{r}%
^{\ast}\mathbf{,}t,r_{s}\mid n)}}{\int d\mathbf{r}^{\prime}n_{T}^{\ast}\left(
\mathbf{r}^{\ast\prime}\right)  e^{\Gamma_{e}(t,r_{s})\Delta u(\mathbf{r}%
^{\ast\prime},,t,r_{s}\mid n)}}. \label{2.15}%
\end{equation}%
\begin{equation}
\Delta u(\mathbf{r}^{\ast}\mathbf{,}t,r_{s}\mid n)=\int d\mathbf{r}^{\prime
}\left(  \overline{c}(\left\vert \mathbf{r}^{\ast}\mathbf{-r}^{\ast\prime
}\right\vert ,t,r_{s})n^{\ast}\left(  \mathbf{r}^{\ast\prime},t,r_{s}\right)
-\overline{c}_{T}(\left\vert \mathbf{r}^{\ast}\mathbf{-r}^{\ast\prime
}\right\vert ,t,r_{s})n_{T}^{\ast}\left(  \mathbf{r}^{\prime},t,r_{s}\right)
\right)  . \label{2.16}%
\end{equation}
Here $n_{T}^{\ast}\left(  \mathbf{r}^{\ast},t,r_{s}\right)  $ is the "trial"
quantum density profile enforcing the associated quantum limit for $n^{\ast
}\left(  \mathbf{r}^{\ast},t,r_{s}\right)  $, and $\overline{c}_{T}(r^{\ast
},t,r_{s})$ is the associated direct correlation function for that limit. See
Appendix \ref{apB} for further details. Equations (\ref{2.15}) and
(\ref{2.16}) are the basis for all the results reported here. Two cases are
considered here, the limit of non-interacting Fermions in a harmonic trap, and
the corresponding system with weak Coulomb interactions.

\section{Classical trap for non-interacting Fermions}

\label{sec3}For a first study of the quantum effects consider an effective
trap whose classical density is the same as the quantum density of
non-interacting Fermions in a harmonic trap. The corresponding trap density in
(\ref{2.15}) and (\ref{2.16}) is denoted by $n_{T}^{\ast}(r^{\ast}%
,t,r_{s})\rightarrow n^{\ast(0)}(r^{\ast},t,r_{s})$ and the direct correlation
function for this case is denoted by $c_{T}(r^{\ast},t,r_{s})\rightarrow
c^{(0)}(r^{\ast},t,r_{s})$. The former is calculated directly from%
\begin{equation}
n^{\ast(0)}\left(  \mathbf{r}^{\ast},t,r_{s}\right)  =2r_{0}^{3}\left\langle
\mathbf{r}\right\vert \left(  e^{\left(  \beta\frac{\widehat{p}^{2}}%
{2m}-\left(  \nu_{0}-\frac{1}{2}m\omega^{2}\widehat{r}^{2}\right)  \right)
}+1\right)  ^{-1}\left\vert \mathbf{r}\right\rangle . \label{3.1}%
\end{equation}
$\left\langle \mathbf{r}\right\vert X\left\vert \mathbf{r}\right\rangle $
denotes a diagonal matrix element in coordinate representation. It has been
assumed that the system is comprised of unpolarized spin $1/2$ particles. A
caret on a variable indicates it is the operator corresponding to that
variable. The parameter $\nu_{0}$ is determined by the condition that the
total average number of particles is the same as the interacting system%
\begin{equation}
\overline{N}(t,r_{s})=2r_{0}^{-3}\int d\mathbf{r}\left\langle \mathbf{r}%
\right\vert \left(  e^{\left(  \beta\frac{\widehat{p}^{2}}{2m}-\left(  \nu
_{0}-\frac{1}{2}m\omega^{2}\widehat{r}^{2}\right)  \right)  }+1\right)
^{-1}\left\vert \mathbf{r}\right\rangle . \label{3.2}%
\end{equation}
Equations (\ref{3.1}) and (\ref{3.2}) can be evaluated in terms of the
harmonic oscillator eigenfunctions and eigenvalues. Instead, here a local
density (Thomas-Fermi) approximation is used. This follows from the
replacement of the operator $\widehat{r}^{2}$ by the corresponding c-number
$r^{2}$. Then the matrix element can be evaluated to give%

\begin{align}
n^{\ast(0)}\left(  \mathbf{r}^{\ast}\right) & \rightarrow\left(  \frac{r_{0}%
}{\lambda}\right)  ^{3}\frac{4}{\sqrt{\pi}}I_{\frac{1}{2}}\left(  \nu
_{0}-\frac{1}{2}\Gamma(t,r_{s})r^{\ast2}\right)
\hspace{0.2in} \\[1.0ex]
\overline{N} &=\left(  \frac{r_{0}}{\lambda}\right)  ^{3}4\pi\int_{0}^{\infty}dr^{\ast
}r^{\ast2}\frac{4}{\sqrt{\pi}}I_{\frac{1}{2}}\left(  \left(  \nu_{0}-\frac
{1}{2}\Gamma(t,r_{s})r^{\ast2}\right)  \right)  . \label{3.3}%
\end{align}
The Fermi function $I_{\alpha}\left(  \beta\mu\right)  $ and thermal de
Broglie wavelength $\lambda$ are given by%
\begin{equation}
I_{\alpha}\left(  \beta\mu\right)  =\int_{0}^{\infty}dx\frac{x^{\alpha}%
}{e^{x-\beta\mu}+1},\hspace{0.2in}\lambda=\left(  \frac{2\pi\hbar^{2}\beta}%
{m}\right)  ^{1/2}. \label{3.5}%
\end{equation}
The validity of this Thomas-Fermi approximation for the conditions considered
here ($\overline{N}=100$) is demonstrated in Appendix \ref{apC}.

The direct correlation function $c^{(0)}(r^{\ast},t,r_{s})$ is non-trivial
because the classical system corresponding to a non-interacting quantum gas
has pairwise interactions needed to reproduce the symmetrization effects. Hence
calculation of properties for this effective classical system is a true
many-body\ problem. The Ornstein-Zernicke equation is used, with the known
exact quantum non-interacting pair correlation function $g^{(0)}(r)$ as input
\cite{DuttaDufty}
\begin{equation}
c^{(0)}\left(  r^{\ast},t,r_{s}\right)  =\left(  g^{(0)}(r^{\ast}%
,t,r_{s})-1\right)  -\overline{n}\int d\mathbf{r}^{\ast\prime}c^{(0)}%
(\left\vert \mathbf{r}^{\ast}\mathbf{-r}^{\ast\prime}\right\vert
,t,r_{s})\left(  g^{(0)}(r^{\ast\prime},t,r_{s})-1\right)  . \label{3.6}%
\end{equation}
Finally, the direct correlation function for the interacting system is
calculated from the coupled HNC and Ornstein-Zernicke equations%

\begin{equation}
\ln\left(  g(r^{\ast},t,r_{s})\right)  =-\phi_{c}^{\ast}(r^{\ast}%
,t,r_{s})+\left(  g(r^{\ast},t,r_{s})-1\right)  -c\left(  r^{\ast}%
,t,r_{s}\right)  , \label{3.7}%
\end{equation}%
\begin{equation}
c\left(  r^{\ast},t,r_{s}\right)  =\left(  g(r^{\ast},t,r_{s})-1\right)
-\overline{n}\int d\mathbf{r}^{\prime}c(\left\vert \mathbf{r}^{\ast
}\mathbf{-r}^{^{\ast}\prime}\right\vert ,t,r_{s})\left(  g(r^{\ast\prime
},t,r_{s})-1\right)  . \label{3.8}%
\end{equation}
Here $\phi_{c}^{\ast}(r^{\ast},t,r_{s})$ is the effective classical pair
interaction representing the uniform electron gas, described in Appendix
\ref{apA}.

\begin{figure}[ptb]
\includegraphics[width=\columnwidth]{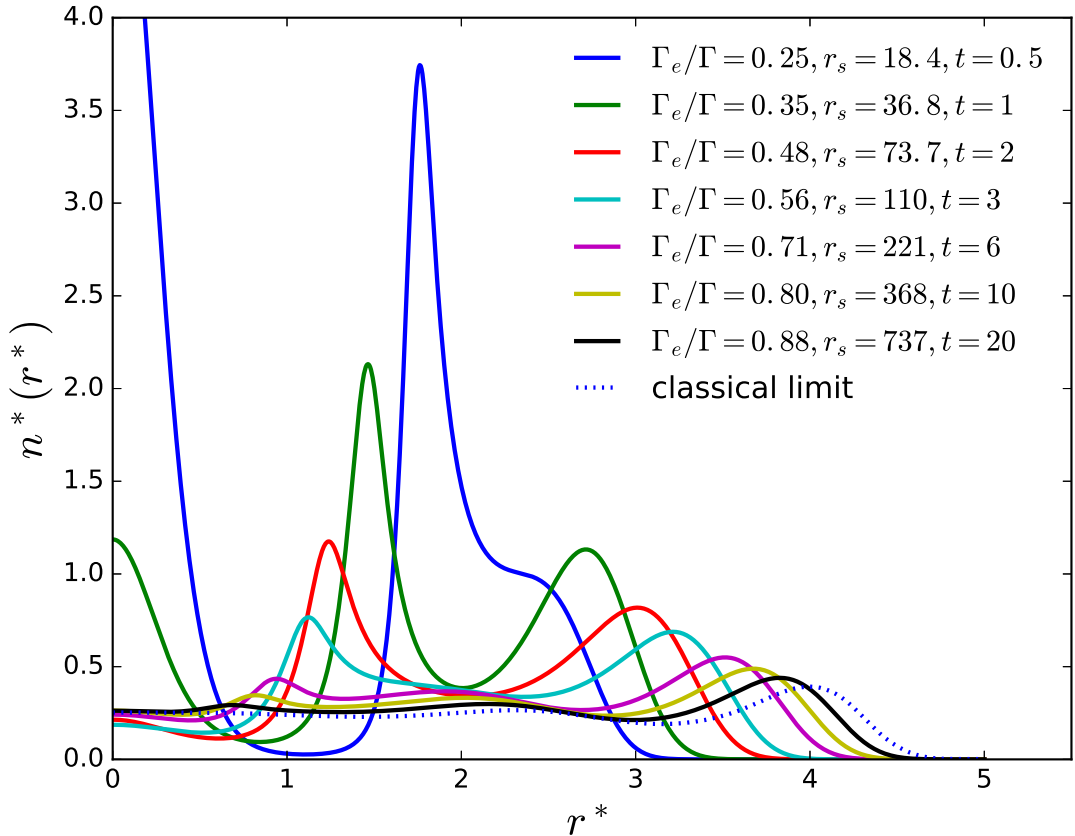}\caption{Onset of quantum effects
for a system of 100 particles. Here $\Gamma=20$ as temperature decreases from
$t=20$ to $t=0.5$.}%
\label{Fig.2}%
\end{figure}

\begin{figure}[ptb]
\includegraphics[width=\columnwidth]{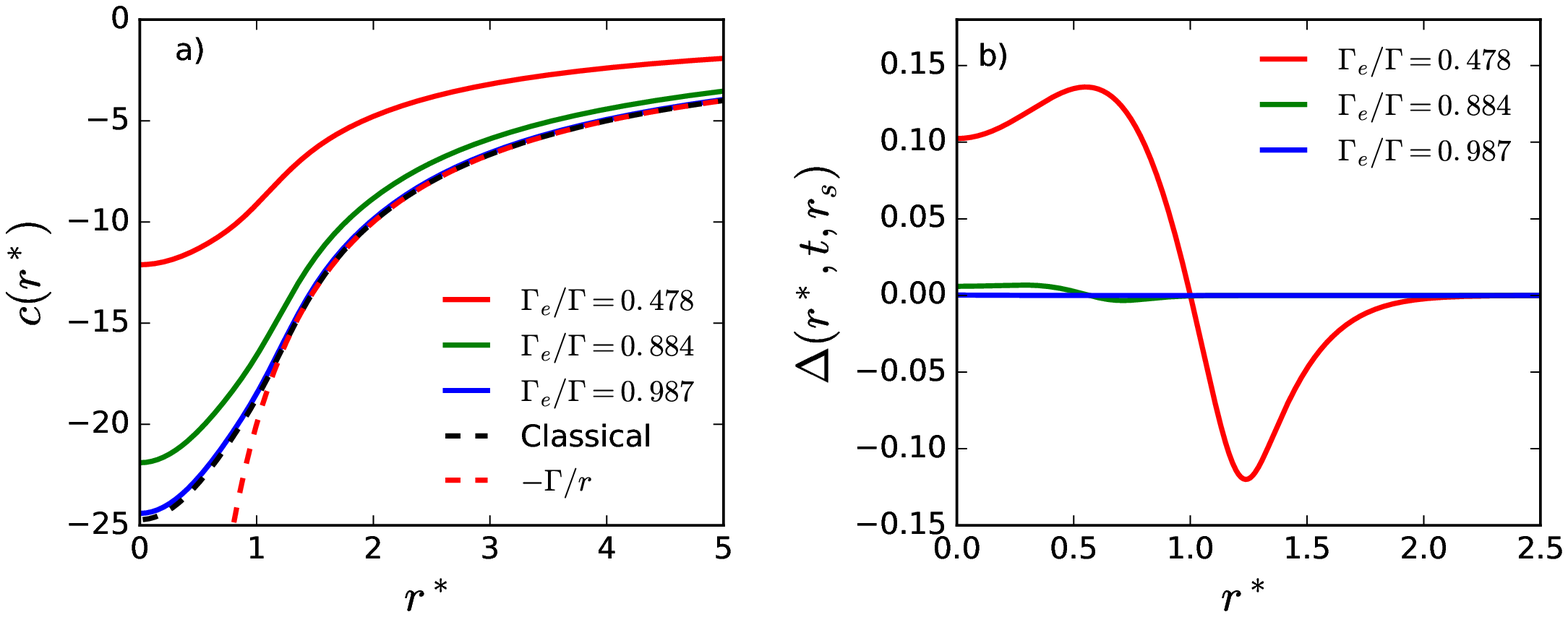} \caption{ Two quantum effects for
$\Gamma=20$. a) Quantum effect on the direct correlation function. The negative of the Coulomb potential is also shown for reference. b) Quantum effect on the shape of the trapping potential near the
origin.}%
\label{Fig.3}%
\end{figure}

Equations (\ref{2.15}) and (\ref{2.16}) for this case are now%
\begin{equation}
n^{\ast}\left(  r^{\ast},t,r_{s}\right)  =N\frac{n^{\ast(0)}\left(  r^{\ast
},t,r_{s}\right)  e^{\Gamma_{e}(t,r_{s})\Delta u(r^{\ast}\mathbf{,}t,r_{s}\mid
n)}}{\int d\mathbf{r}^{\prime}n^{\ast(0)}\left(  r^{\ast\prime}\right)
e^{\Gamma_{e}(t,r_{s})\Delta u(\mathbf{r}^{\ast\prime},,t,r_{s}\mid n)}},
\label{3.9}%
\end{equation}%
\begin{equation}
\Delta u(\mathbf{r}^{\ast}\mathbf{,}t,r_{s}\mid n)=\int d\mathbf{r}^{\prime
}\left(  \overline{c}(\left\vert \mathbf{r}^{\ast}\mathbf{-r}^{\ast\prime
}\right\vert ,t,r_{s})n^{\ast}\left(  r^{\ast\prime},t,r_{s}\right)
-\overline{c}^{(0)}(\left\vert \mathbf{r}^{\ast}\mathbf{-r}^{\ast\prime
}\right\vert ,t,r_{s})n^{\ast(0)}\left(  r^{\ast\prime},t,r_{s}\right)
\right)  . \label{3.10}%
\end{equation}
The quantum input for this classical description is two-fold. The first is a
modification of the Coulomb interactions among charges via $\phi_{c}^{\ast
}(r^{\ast},t,r_{s})$, due to both diffraction and exchange effects. These
occur through the direct correlations $\overline{c}(r^{\ast},t,r_{s})$.
Additional quantum effects occur due to the modification of the shape and
intensity of the harmonic trap. These occur through $n^{\ast(0)}\left(
r^{\ast},t,r_{s}\right)  $. To explore these a series of density profiles \ is
shown in Figure \ref{Fig.2} for values of $t,r_{s\text{ }}$corresponding to
the line $\Gamma=20$ in Figure \ref{Fig.1}. Without quantum effects all
profiles would be the same as the classical limit shown. The observed
classical shell structure in that case is due entirely to strong Coulomb coupling with no
quantum effects. As the values of $t,r_{s\text{ }}$ are decreased this Coulomb
shell is distorted and shifted inward, corresponding to a weakening of the
Coulomb repulsion through a decreasing effective coupling $\Gamma_{e}%
(t,r_{s})$. This weakening of Coulomb correlations in $c(r^{\ast},t,r_{s})$ is
displayed in Figure \ref{Fig.3}a. The direct correlation function has quantum
effects that enter the HNC theory only through the effective pair potential
(Appendix \ref{apA}). The latter has a Coulomb tail whose amplitude is
decreased by $\Gamma_{e}/\Gamma$ so that long range correlations are weakened.
At shorter distances the Coulomb singularity is removed in the effective pair
potential due to diffraction effects. The classical direct correlation
function is finite at $\mathbf{r}^{\ast}=0$ for sufficiently strong coupling
due to Coulomb correlations in spite of the singular Coulomb potential.
However, with quantum diffraction effects the effective pair potential is
non-singular and the direct correlation function remains finite $\mathbf{r}%
^{\ast}=0$ even at weak coupling. These qualitative changes are illustrated
for three cases in Figure \ref{Fig.3}a corresponding to $t=200,20,$ and $2$ in
Figure \ref{Fig.2}. The smaller values at $\mathbf{r}^{\ast}=0$ tend to
enhance shell formation while the weaker coupling of $\Gamma_{e}/\Gamma$ tends
to decrease it.

\begin{figure}[ptb]
\includegraphics[width=\columnwidth]{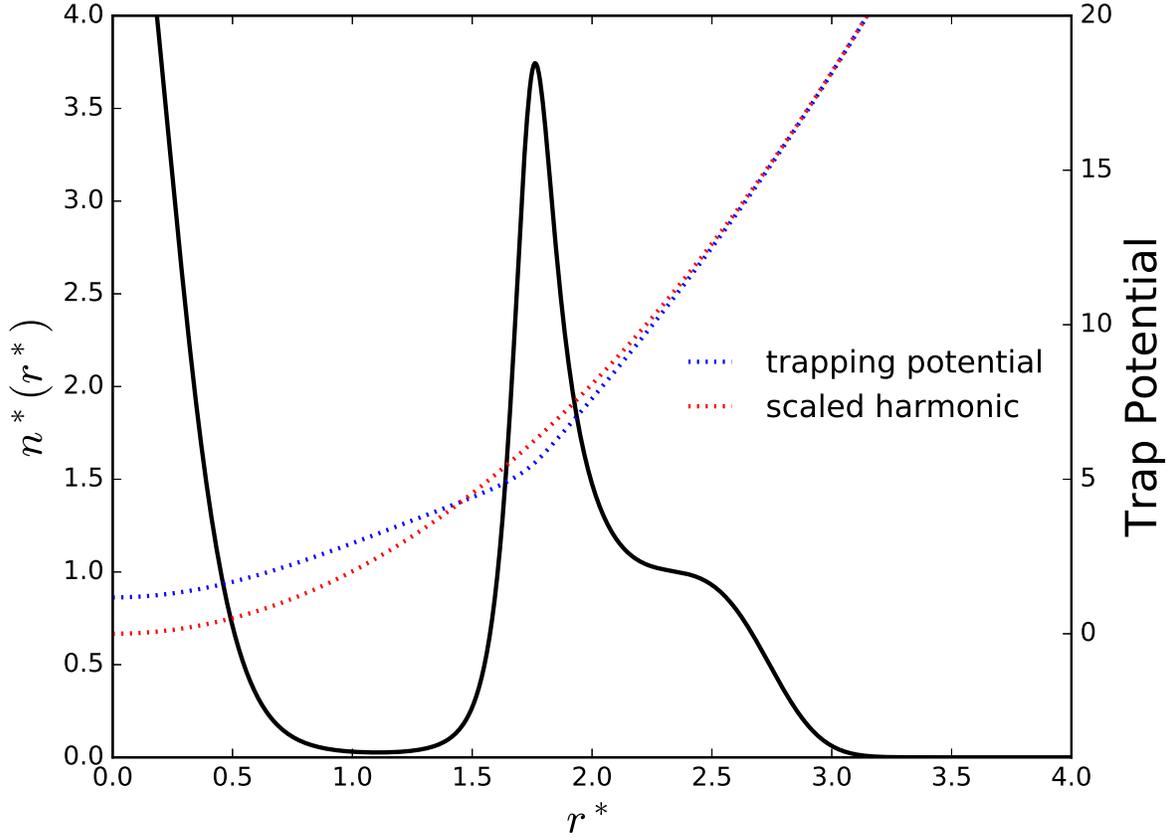} \caption{ The low temperature
quantum effect of the trapping potential on the density $n(r)$. The system is
strongly coupled ($\Gamma=20$) with $r_{s}=18.4$ and $t=0.5$. The scaled harmonic function is shown, as well
as the full trapping potential. }%
\label{Fig.4}%
\end{figure}

A qualitatively new consequence of quantum effects occurs at the lowest value
of $t=0.5$ and $r_s=18.4$. A strong single shell occurs that is
 unrelated to the classical Coulomb shell structure and is due entirely to a
change in shape of the confining potential. To be more explicit, write the
confining potential, or equivalently $\overline{\nu}_{c}(\mathbf{r}^{\ast
},t,r_{s})$, as%
\begin{equation}
\overline{\nu}_{c}^{(0)}(\mathbf{r}^{\ast},t,r_{s})-\overline{\nu}_{c}%
^{(0)}(\mathbf{0},t,r_{s})=\frac{\Gamma}{\Gamma_{e}}\frac{1}{2}\mathbf{r}%
^{\ast2}+\Delta\left(  \mathbf{r}^{\ast},t,r_{s}\right)  . \label{3.10a}%
\end{equation}
There are two quantum effects evident in this form, an increase in amplitude
of the harmonic potential by $\Gamma/\Gamma_{e}$, and a change in shape
represented by $\Delta\left(  \mathbf{r}^{\ast},t,r_{s}\right)  $. The change
in amplitude of the harmonic potential is a reflection of its enhancement
relative to $\overline{c}(r^{\ast},t,r_{s})$ and is largely responsible for
the increased confinement observed in all density profiles of Figure
\ref{Fig.2}. As the shells are pulled inwards, this also tends to cause a
population transfer to the outer shell. However, at the lowest temperatures
the change in shape from the harmonic form becomes large. It is this
distortion that is responsible for the onset of the new shell structure seen in
Figure \ref{Fig.2}. This is confirmed in Figure \ref{Fig.4} which shows the
superposition of the shell and the local distortion of the confining potential
relative to its harmonic form. The origin of this distortion is the Fermi
statistics of the non-interacting particles which force the trap density to go
to zero at a finite radius as $t\rightarrow0$ (Appendix \ref{apB}). This
translates into a hard wall for the effective confining potential, and an
associated shell structure (even in a classical fluid hard wall confinement
leads to shell structure). The predicted location of the $t=0$ wall in
Appendix \ref{apB} is $1.77$, very close to that observed in Figure
\ref{Fig.4} at $t=0.5$.

\section{Classical trap with weak Coulomb interactions}

\label{sec4}Now consider the same analysis based on (\ref{2.15}) and
(\ref{2.16}), but with a better choice for the effective confining potential
to include some effects of the Coulomb interactions on the classical confining
potential. \ This change does not affect $\overline{c}(r^{\ast},t,r_{s})$,
which is the same as in the previous section. The new choice is defined by
imposing a weak coupling limit for which the corresponding trap density is obtained from a
quantum density functional calculation including Hartree and exchange
interactions in a local density approximation, $n_{T}^{\ast}\left(
\mathbf{r}^{\ast},t,r_{s}\right)  \rightarrow n^{\ast\left(  w\right)
}\left(  \mathbf{r}^{\ast},t,r_{s}\right)  $ given by (\ref{b.13}). The
details are discussed in Appendix \ref{apB2}. Accordingly, the corresponding
classical limit for the trial direct correlation function is its weak coupling
expansion to first order in $\Gamma$, $\overline{c}_{T}(r^{\ast}%
,t,r_{s})\rightarrow\overline{c}^{(0)}(r^{\ast},t,r_{s})+\Gamma\overline
{c}^{(1)}(r^{\ast},t,r_{s})$, and (\ref{2.15}) and (\ref{2.16}) become
\begin{equation}
n^{\ast}\left(  \mathbf{r}^{\ast},t,r_{s}\right)  =\overline{N}\frac
{n^{\ast\left(  w\right)  }\left(  \mathbf{r}^{\ast},t,r_{s}\right)
e^{\Gamma_{e}(t,r_{s})\Delta u(\mathbf{r}^{\ast}\mathbf{,}t,r_{s}\mid n)}%
}{\int d\mathbf{r}^{\prime}n^{\ast\left(  w\right)  }\left(  \mathbf{r}%
^{\ast\prime}\right)  e^{\Gamma_{e}(t,r_{s})\Delta u(\mathbf{r}^{\ast\prime
},,t,r_{s}\mid n)}}. \label{4.1}%
\end{equation}
\begin{align}
\Delta u(\mathbf{r}^{\ast}\mathbf{,}t,r_{s}  &  \mid n)=\int d\mathbf{r}%
^{\prime}\left(  \overline{c}(\left\vert \mathbf{r}^{\ast}\mathbf{-r}%
^{\ast\prime}\right\vert ,t,r_{s})n^{\ast}\left(  \mathbf{r}^{\ast\prime
},t,r_{s}\right)  \right. \nonumber\\
&  \left.  -\left(  \overline{c}^{(0)}(\left\vert \mathbf{r}^{\ast}%
\mathbf{-r}^{\ast\prime}\right\vert ,t,r_{s})+\Gamma\overline{c}%
^{(1)}(\left\vert \mathbf{r}^{\ast}\mathbf{-r}^{\ast\prime}\right\vert
,t,r_{s})\right)  n^{\ast\left(  w\right)  }\left(  \mathbf{r}^{\prime
},t,r_{s}\right)  \right)  \label{4.2}%
\end{align}
The direct correlation functions $\overline{c}(r^{\ast},t,r_{s})$ and
$\overline{c}^{(0)}(r^{\ast},t,r_{s})$ are again calculated in the HNC
approximation using (\ref{3.6}) - (\ref{3.8}). Also, the weak coupling
coefficient $\overline{c}^{(1)}(r^{\ast},t,r_{s})$ is obtained numerically
from these equations for asymptotically small $\Gamma$.

\begin{figure}[ptb]
\includegraphics[width=\columnwidth]{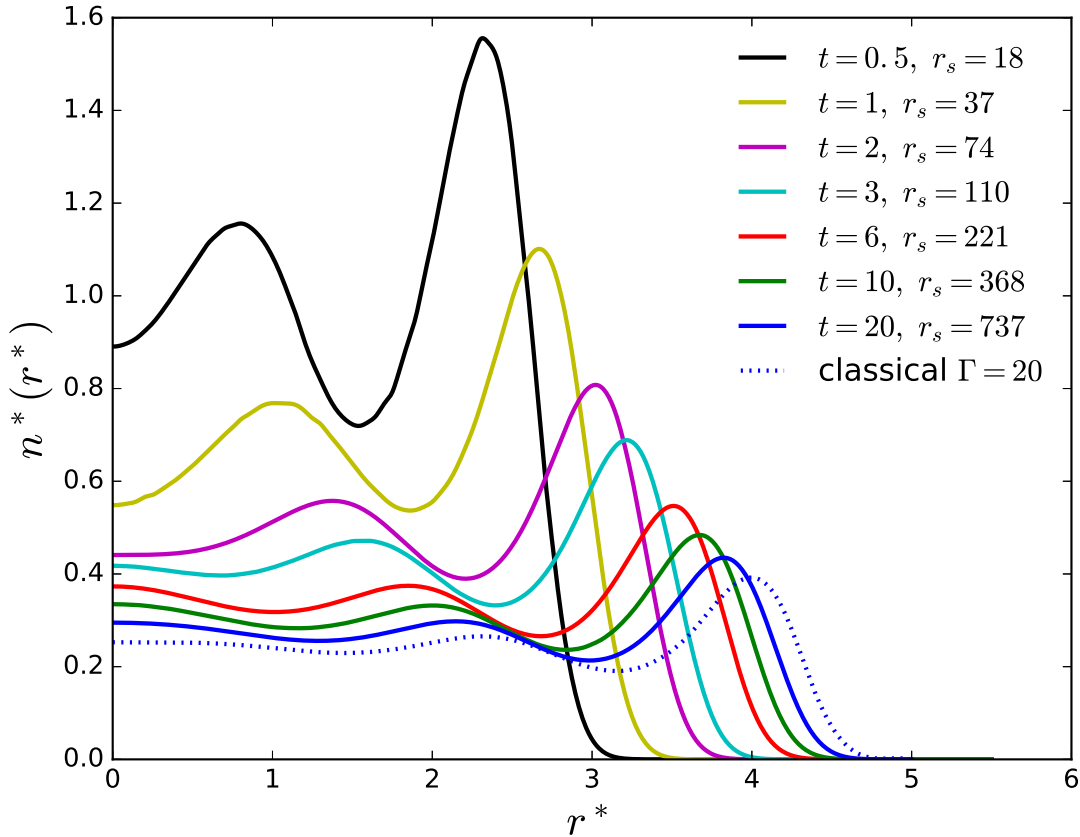}\caption{Onset of quantum effects
for a system of 100 particles. Here $\Gamma=20$ as temperature decreases from
$t=20$ to $t=0.5$.}%
\label{Fig.5}%
\end{figure}

\begin{figure}[ptb]
\includegraphics[width=\columnwidth]{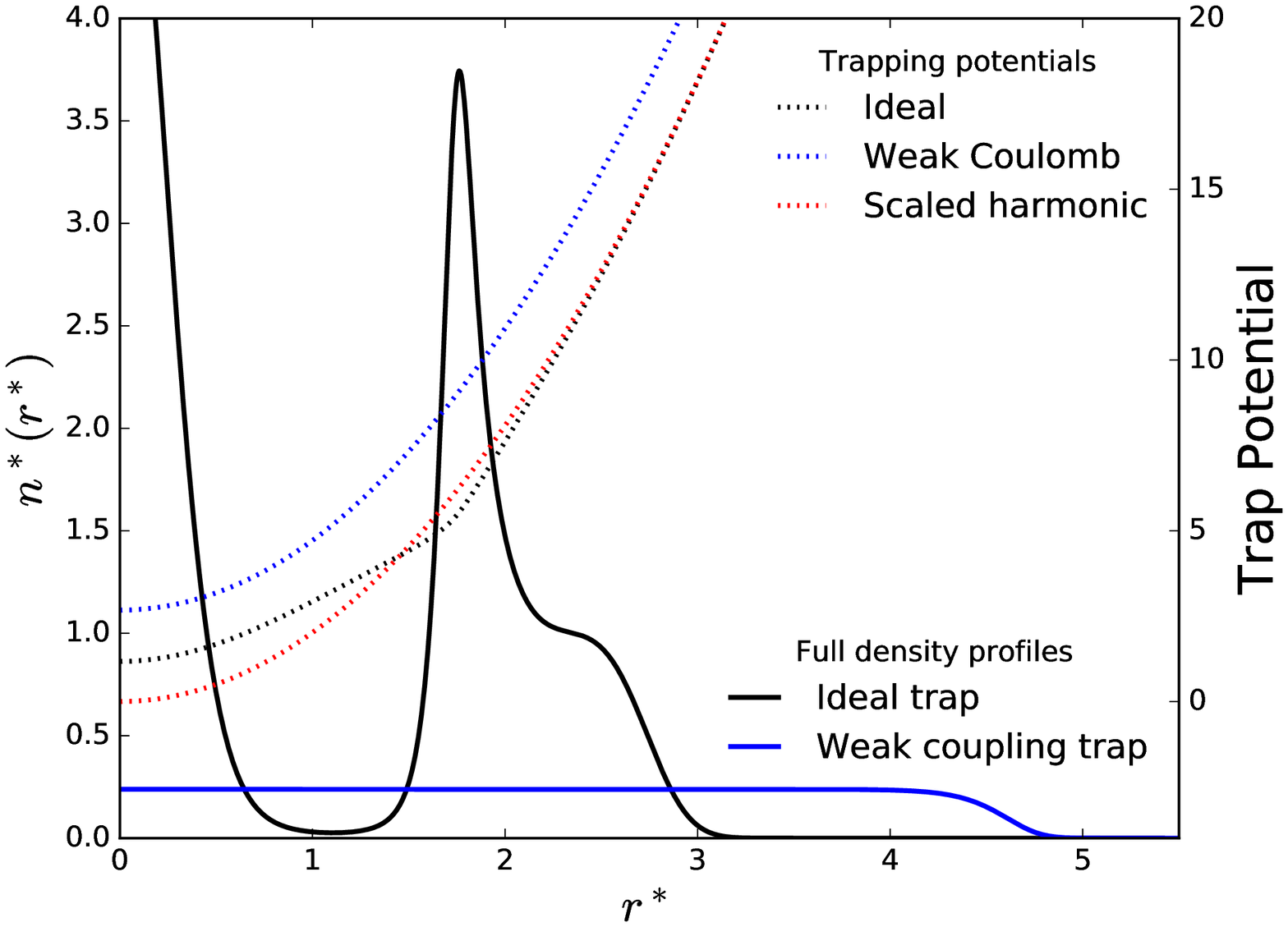}\caption{Effect of including weak Coulomb interactions on trapping potential distortion.  Black dashed line is the distorted trapping potential for non-interacting particles.  Blue dashed line is the distorted trapping potential when weak Coulomb interactions are included. For reference, the red dashed line shows the scaled harmonic potential $\frac{1}{2}\frac{\Gamma}{\Gamma_e}r^{*2}$ with no shape distortion.  Here $t=0.5$ and $\Gamma=20$.}%
\label{Fig.6}%
\end{figure}

Figure \ref{Fig.5} shows the density profiles for the same temperatures as in
Figure \ref{Fig.2} along the line $\Gamma=20$ in Figure \ref{Fig.1}. The
results are quite similar at the high temperatures, e.g. $t=20$, as the
classical limit is approached. However, at all lower temperatures there is a
qualitative difference between Figures \ref{Fig.5} and \ref{Fig.2}. In the
latter case the intermediate peak diminishes and the new shell at small
$r^{\ast}$ grows as the temperature decreases until a single dominant peak is
formed at the lowest temperature. In contrast, the outer and intermediate
peaks of Figure \ref{Fig.5} change in a unified fashion as the overall density
profile contracts with decreasing temperature. The two peak structure is
maintained with only quantitative changes occurring due to quantum effects - no
new shell structure is seen as in Figure \ref{Fig.2}. As indicated in (\ref{3.10a}), the
quantum effects on the confining potential are an enhancement of the harmonic
form and a distortion of that form. The distortion $\Delta\left(
\mathbf{r}^{\ast},t,r_{s}\right)  $ is now very much decreased by the
inclusion of weak Coulomb interactions in the determination of the classical
confining potential, eliminating the new "hard wall" shell structure of Figure
\ref{Fig.2}. This is illustrated in Figure \ref{Fig.6} for $t=0.5$.

\begin{figure}[ptb]
\includegraphics[width=\columnwidth]{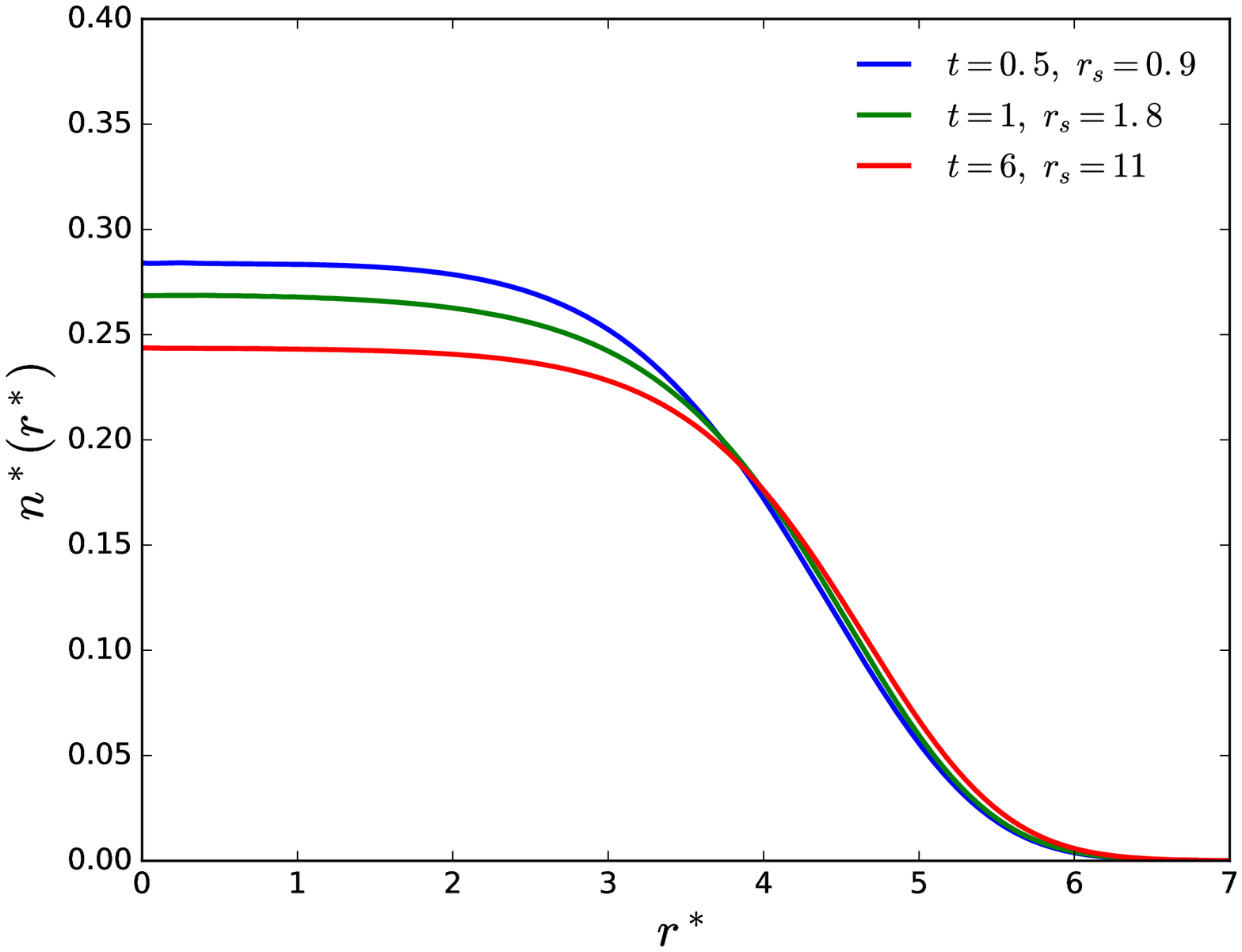}\caption{Density profiles for low temperatures ($t=0.5, 1, 6$) at $\Gamma=1$ for weak Coulomb interactions.  No shell structure is formed at small temperatures for the weakly coupled case. }%
\label{Fig.7}%
\end{figure}

The quantum effects on the amplitude and location of the shells in Figure \ref{Fig.5}
are quite significant. For example, at $t=1$ the outer peak increases by a factor of
2.8 relative to the classical value. The contraction is largely due
to the factor $\Gamma(t,r_{s})/\Gamma_{e}(t,r_{s})$ which changes from $1.13$
at $t=20$ to $2.86$ at $t=1$. The results discussed thus far are all for the
strong coupling condition $\Gamma=20$. This was chosen because shell structure
is present for these conditions even in the classical limit. It is instructive
now to consider the case $\Gamma=1$ for which there is no classical shell
structure. Figure \ref{Fig.7} shows the results for $t=6,1,$ and $0.5$. In
contrast to the strong coupling case, $t=6$ is very close to the classical
limit. The contraction of the profile is the dominant quantum effect at lower
temperatures, and there is no shell structure evident in any case.

\section{Discussion}

\label{sec5}The classical shell structure for strong coupling conditions in
the upper right corner of Figure \ref{Fig.1} has provided a wealth of insight
into formation of shell structure due to Coulomb correlations. Here these
studies have been extended in the direction of additional quantum effects. The
method chosen, an equivalent quantum system, allows inclusion of the diverse
classical effects into an extension via effective pair potentials and
effective confinement potentials. The quantum effects are included in the
modification of these potentials from their classical Coulomb and harmonic
forms in a controlled way defined by the formalism of references
\cite{DuftyDutta,DuttaDufty}. Two approximate implementations of that
formalism have been described. In both, the pair correlations among charges
expressed by the direct correlation function $\overline{c}(r^{\ast},t,r_{s})$
are calculated from the classical HNC liquid state theory, known to be
accurate for strong correlations, e.g. $\Gamma=20.$ The qualitative effects of
quantum mechanics are illustrated in Figure \ref{Fig.3}a. The first
approximation for the effective confining potential is that which gives the
exact quantum density profile for non-interacting charges. The result is a
scaling of the original harmonic trap by a factor $\Gamma(t,r_{s})/\Gamma
_{e}(t,r_{s})$ which tends to increase the confinement relative to the Coulomb
correlations. In addition there is a distortion of the harmonic form at low
temperatures that produces a "hard wall" associated with the vanishing of the
non-interacting density at a finite value of $r^{\ast}$. This leads to a new
shell structure not related to Coulomb correlations.

The second choice for the confining potential, described in Section
\ref{sec4}, is that which gives the density profile for a weak coupling
quantum density functional calculation. This potential includes the effects of
Coulomb interactions. It has a similar scaling of the harmonic form, but no
longer shows the strong distortion (compare Figures \ref{Fig.3}b and
\ref{Fig.6}) and hence no new shell structure. In fact the profiles of Figure
\ref{Fig.5} at $\Gamma=20$ appear like a self-similar contraction constrained
by the normalization to $\overline{N}=100$. The choice of parameters
$\Gamma=20,\overline{N}=100$ was made to insure multiple shells in the
reference classical limit. The brief consideration of $\Gamma=1,\overline
{N}=100$ in Figure \ref{Fig.7} confirms that there is no new shell structure
induced solely by quantum effects.

Clearly there is more to be done with this classical description of a quantum
system, such as $t<0.5$ and much smaller $\overline{N}$ to make direct
connection with the literature on quantum dots. Presumably, for such conditions
the local density approximation will need to be relaxed. A different direction
for application is the replacement of the harmonic trap by a Coulomb potential
to calculate the electron distribution about an ion. This is the first step in
addressing the more practical case of determining the electronic configuration
in a distribution of ionic sources. Such configurations are required to
compute the forces in quantum molecular dynamics simulations for the ions in
warm, dense matter at finite temperatures where traditional density functional
methods fail \cite{WDM}.

\section{Acknowledgements}

The authors are indebted to Michael Bonitz for his comments on an earlier draft. 
This research has been supported in part by NSF/DOE Partnership in Basic
Plasma Science and Engineering award DE-FG02-07ER54946 and by US DOE Grant
DE-SC0002139. 

\appendix

\section{Effective Classical Direct Correlation Function}

\label{apA}The density profile for charges in a trap is governed by both the
confining potential and the correlations among the particles in the trap. The
latter appear in (\ref{2.16}) via the direct correlation function $c(r,\mu
_{c},\beta_{c})=\Gamma_{e}(t,r_{s})\overline{c}(r^{\ast},t,r_{s})$. In this
appendix, the approximate evaluation of these correlations from the HNC
integral equations of liquid state theory \cite{Hansen} using an effective
pair potential is summarized.

As noted in Section \ref{sec2}, the correlations for the non-uniform charges
in the trap are approximated by those for a uniform electron gas. The
calculation of these correlations from an effective classical system has been
described in some detail elsewhere \cite{DuftyDutta}, so only the relevant
equations are reproduced here for completeness. The approximate effective pair
potential\ used there is%

\begin{equation}
\phi_{c}^{\ast}(r^{\ast},t,r_{s})=\beta_{c}\phi_{c}(r)=\phi_{c}^{\ast
(0)}(r^{\ast},t,r_{s})+\frac{1}{n}\int\frac{d\mathbf{k}}{\left(  2\pi\right)
^{3}}e^{-i\mathbf{k\cdot r}}\left(  \frac{1}{S^{RPA}(k)}-\frac{1}{S^{(0)}%
(k)}\right)  . \label{a.1}%
\end{equation}
Here $S^{RPA}(k)$ and $S^{(0)}(k)$ are the static structure factor for the
random phase approximation and ideal gas, respectively. The first term
$\left(  \beta_{c}\phi_{c}(r)\right)  ^{(0)}$ is the effective potential for
the ideal quantum gas obtained by inverting the coupled ideal gas HNC
equations \cite{Hansen}, i.e. eqs. (\ref{3.7}) and (\ref{3.8}) specialized to
the ideal gas%
\begin{equation}
\ln\left(  g^{(0)}(r^{\ast},t,r_{s})\right)  =-\phi_{c}^{\ast(0)}(r^{\ast
},t,r_{s})+\left(  g^{(0)}(r^{\ast},t,r_{s})-1\right)  -c^{(0)}\left(
r^{\ast},t,r_{s}\right)  , \label{a.2}%
\end{equation}%
\begin{equation}
c^{(0)}\left(  r^{\ast},t,r_{s}\right)  =\left(  g^{(0)}(r^{\ast}%
,t,r_{s})-1\right)  -\overline{n}\int d\mathbf{r}^{\prime}c^{(0)}(\left\vert
\mathbf{r}^{\ast}\mathbf{-r}^{^{\ast}\prime}\right\vert ,t,r_{s})\left(
g^{(0)}(r^{\ast\prime},t,r_{s})-1\right)  , \label{a.3}%
\end{equation}
using the known exact ideal gas pair correlation function for $g^{(0)}%
(r,t,r_{s})$. Finally, with $\phi_{c}^{\ast}(r^{\ast},t,r_{s})$ determined in
this way the direct correlation function for the interacting system is
calculated from the full coupled HNC equations (\ref{3.7}) and (\ref{3.8}).

As a practical matter, a simplified representation of (\ref{a.1}) has been
proposed \cite{DD13}. The ideal gas contribution $\phi_{c}^{\ast\left(
0\right)  }(r^{\ast},t,r_{s})$ is the same, but the contribution from the
Coulomb interactions is modeled by the exact low density, weak coupling
functional form first derived by Kelbg \cite{Kelbg}. Here that form is
parameterized to include the exact low density value for the pair correlation
function at $r=0$ \cite{Filinov04}, and the large $r$ behavior of the more complete form
(\ref{a.1})%
\begin{equation}
\phi_{c}^{\ast}(r^{\ast},t,r_{s})\simeq\phi_{c}^{\ast(0)}(r^{\ast}%
,t,r_{s})+\Delta_{K}^{\ast}\left(  r^{\ast},\Gamma_{e},r_{s}\right)  ,
\label{a.4}%
\end{equation}
with%
\begin{equation}
\Delta_{K}^{\ast}\left(  r^{\ast},\Gamma_{e},r_{s}\right)  \equiv\frac
{\Gamma_{e}}{r^{\ast}}\left(  1-\exp(-\left(  ar^{\ast}\right)  ^{2}%
)+\sqrt{\pi}\frac{ar^{\ast}}{\gamma}\operatorname{erfc}(\gamma
ar^{\ast})\right)  . \label{a.5}%
\end{equation}
Here%
\begin{equation}
a=\left(  r_{s}/\Gamma_{e}\right)  ^{1/2},\hspace{0.2in}\gamma\left(
\Gamma_{e}r_{s}\right)  =-\frac{\left(  \pi\Gamma_{e}r_{s}\right)  ^{1/2}%
}{\mathrm{\ln}s(\Gamma_{e}r_{s})}, \label{a.6}%
\end{equation}
and $s(\Gamma_{e}r_{s})$ is the two electron relative coordinate Slater sum at
$r^{\ast}=0$%
\begin{equation}
s(\Gamma_{e}r_{s})=-4\left(  \pi\Gamma_{e}r_{s}\right)  ^{1/2}\int_{0}%
^{\infty}dye^{-y^{2}}\frac{y}{1-e^{\pi\left(  \Gamma_{e}r_{s}\right)
^{1/2}/y}}. \label{a.7}%
\end{equation}
Also $\Gamma_{e}$ is the effective coupling constant of (\ref{2.12}). Clearly,
(\ref{a.4}) has the computational advantage that $\Delta_{K}^{\ast}\left(
r^{\ast},\Gamma_{e},r_{s}\right)  $ is an explicit, analytic function of the
input parameters $t,r_{s}$. The results obtained for correlations using
(\ref{a.4}) are quite similar to those obtained using (\ref{a.1}).

\section{Effective Classical Trap Potential}

\label{apB}The effective classical description of the local density for
charges confined in a harmonic trap is given by \cite{DuftyDutta,DuttaDufty}
\begin{equation}
\ln\left(  n\left(  \mathbf{r}\right)  \lambda_{c}^{3}\right)  =\left(
\beta_{c}\mu_{ec}-\beta_{c}v_{c}(\mathbf{r})\right)  +\int d\mathbf{r}%
^{\prime}c(\left\vert \mathbf{r-r}^{\prime}\right\vert ,\mu_{c},\beta
_{c})n\left(  \mathbf{r}^{\prime}\right)  . \label{b.1}%
\end{equation}
where $n\left(  \mathbf{r}\right)  $ is the desired charge density and
$c(\left\vert \mathbf{r-r}^{\prime}\right\vert ,\mu_{c},\beta_{c})$ is the
direct correlation function for the homogeneous electron gas calculated as
described in Appendix \ref{apA}. To complete the description it is necessary
to choose the effective trap potential and chemical potential, i.e. $\left(
\beta_{c}\mu_{ec}-\beta_{c}v_{c}(\mathbf{r})\right)  .$ This is done by
requiring that the effective trap reproduce a chosen approximate quantum
density valid in some limit. In this way, some limiting quantum information is
provided via the effective trap.

It is useful to express (\ref{b.2}) in the equivalent form (\ref{2.6}) that
includes the normalization explicitly%
\begin{equation}
n\left(  \mathbf{r},\mu_{c},\beta_{c}\right)  =\overline{N}\frac
{e^{-U(\mathbf{r},\mu_{c},\beta_{c})}}{\int d\mathbf{r}^{\prime}%
e^{-U(\mathbf{r}^{\prime},\mu_{c},\beta_{c})}}, \label{b.2}%
\end{equation}%
\begin{equation}
U(\mathbf{r,}\mu_{c},\beta_{c})=-\nu_{c}(\mathbf{r,}\mu_{c},\beta_{c}%
)-\frac{\overline{N}}{\int d\mathbf{r}^{\prime}e^{-U(\mathbf{r}^{\prime}%
,\mu_{c},\beta_{c})}}\int d\mathbf{r}^{\prime}e^{-U(\mathbf{r}^{\prime}%
,\mu_{c},\beta_{c})}c(\left\vert \mathbf{r-r}^{\prime}\right\vert ,\mu
_{c},\beta_{c}). \label{b.3}%
\end{equation}
Recall the notation that $\nu_{c}(\mathbf{r,}\mu_{c},\beta_{c})=\beta_{c}%
\mu_{c}(\mathbf{r})=$ $\beta_{c}\mu_{ec}-\beta_{c}v_{c}(\mathbf{r})$.

Let $\left(  \beta_{c}\mu_{ec}-\beta_{c}v_{c}(\mathbf{r})\right)  _{T}$ denote
the effective trap potential and chemical potential in some chosen limit. The
density profile in that limit, $n_{T}\left(  \mathbf{r},\mu_{c},\beta
_{c}\right)  ,$ is therefore
\begin{equation}
\ln\left(  n_{T}\left(  \mathbf{r},\mu_{c},\beta_{c}\right)  \lambda_{c}%
^{3}\right)  =\left(  \beta_{c}\mu_{ec}-\beta_{c}v_{c}(\mathbf{r})\right)
_{T}+\int d\mathbf{r}^{\prime}c_{T}(\left\vert \mathbf{r-r}^{\prime
}\right\vert ,\mu_{c},\beta_{c})n_{T}\left(  \mathbf{r}^{\prime},\mu_{c}%
,\beta_{c}\right)  . \label{b.4}%
\end{equation}
Here $c_{T}(r,\mu_{c},\beta_{c})$ is the direct correlation function
corresponding in the classical form to the quantum limit considered.  The limit must be such that an independent quantum calculation of $n_{T}\left(
\mathbf{r},\mu_{c},\beta_{c}\right)  $ can be implemented practically, and the
corresponding $c_{T}(r,\mu_{c},\beta_{c})$ can be identified. Then with
$c_{T}(r,\mu_{c},\beta_{c})$ and $n_{T}\left(  \mathbf{r},\mu_{c},\beta
_{c}\right)  $ known, equation (\ref{b.4}) defines the effective classical
trap that gives the exact quantum density in the limit considered. The choice
for the approximate effective trap in (\ref{b.1}) is now made as%
\begin{equation}
\left(  \beta_{c}\mu_{ec}-\beta_{c}v_{c}(\mathbf{r})\right)  \rightarrow
\left(  \beta_{c}\mu_{ec}-\beta_{c}v_{c}(\mathbf{r})\right)  _{T}. \label{b.5}%
\end{equation}
This assures the exact behavior $n_{T}\left(  \mathbf{r},\mu_{c},\beta
_{c}\right)  $ is recovered in the appropriate limit. With this choice
(\ref{b.2}) and (\ref{b.3}) become
\begin{equation}
n\left(  \mathbf{r},\mu_{c},\beta_{c}\right)  =N\frac{n_{T}\left(
\mathbf{r},\mu_{c},\beta_{c}\right)  e^{\Delta U(\mathbf{r},\mu_{c},\beta
_{c}\mid n)}}{\int d\mathbf{r}^{\prime}n_{T}\left(  \mathbf{r}^{\prime
}\right)  e^{\Delta U(\mathbf{r}^{\prime},\mu_{c},\beta_{c}\mid n)}}.
\label{b.6}%
\end{equation}%
\begin{equation}
\Delta U(\mathbf{r},\mu_{c},\beta_{c}\mid n)=\int d\mathbf{r}^{\prime}\left(
c(\left\vert \mathbf{r-r}^{\prime}\right\vert ,\mu_{c},\beta_{c})n\left(
\mathbf{r}^{\prime},\mu_{c},\beta_{c}\right)  -c_{T}(\left\vert \mathbf{r-r}%
^{\prime}\right\vert ,\mu_{c},\beta_{c})n_{T}\left(  \mathbf{r}^{\prime}%
,\mu_{c},\beta_{c}\right)  \right)  . \label{b.7}%
\end{equation}
Here it has been required that $\int d\mathbf{r}n_{T}\left(  \mathbf{r}%
,\mu_{c},\beta_{c}\right)  =\overline{N}$. Equations (\ref{2.15}) and
(\ref{2.16}) are the dimensionless forms of (\ref{b.6}) and (\ref{b.7}) quoted
in the text.

\subsection{Non-interacting charges limit}

\label{apB1}The simplest choice for an imposed limit by the confining
potential is that for non-interacting charges in a harmonic trap. This choice
properly includes the non-classical effects of exchange symmetry. The density
in this case $n_{T}^{\ast}(r^{\ast},t,r_{s})\rightarrow n^{\ast(0)}(r^{\ast
},t,r_{s})$ is given by the matrix element in (\ref{3.1}), which can be
evaluated directly as a sum over eigenfunctions $\psi_{\alpha}\left(
\mathbf{r}\right)  $ and eigenvalues $\epsilon_{\alpha}$ of the harmonic
oscillator Hamiltonian%
\begin{equation}
n_{T}^{(0)}\left(  \mathbf{r},\mu_{c},\beta_{c}\right)  =\sum_{\alpha
}\left\vert \psi_{\alpha}\left(  \mathbf{r}\right)  \right\vert ^{2}\left(
e^{\left(  \beta\epsilon_{\alpha}-\nu_{0}\right)  }+1\right)  ^{-1}.
\label{b.8}%
\end{equation}
The activity $\nu_{0}$ is determined by the condition that the density
integrate to $\overline{N}$. A simpler practical approximation is given by the
Thomas-Fermi or local density approximation%
\begin{align}
n^{(0)}(\mathbf{r},\mu_{c},\beta_{c})  &  \simeq\frac{2}{h^{3}}\int
d\mathbf{p}\left(  e^{-\nu_{0}}e^{\beta\left(  \frac{p^{2}}{2m}+v(r)\right)
}+1\right)  ^{-1}\nonumber\\
&  =\lambda^{-3}\frac{4}{\sqrt{\pi}}I_{\frac{1}{2}}(\nu_{0}-\beta
v(\mathbf{r})) \label{b.9}%
\end{align}
where $v(r)$ is the harmonic trap potential, and the Fermi function
$I_{\alpha}\left(  \nu_{0}\right)  $ and thermal de Broglie wavelength
$\lambda$ are defined by
\begin{equation}
I_{\alpha}\left(  \nu_{0}\right)  =\int_{0}^{\infty}dx\frac{x^{\alpha}%
}{e^{x-\nu_{0}}+1},\hspace{0.2in}\lambda=\left(  \frac{2\pi\hbar^{2}\beta}%
{m}\right)  ^{1/2}. \label{b.9a}%
\end{equation}
The validity of this Thomas-Fermi approximation for the conditions considered
here is demonstrated in Appendix \ref{apC}.

With this choice for the reference density (\ref{b.6}) and (\ref{b.7})
becomes
\begin{equation}
n\left(  \mathbf{r},\mu_{c},\beta_{c}\right)  =N\frac{n^{(0)}\left(
\mathbf{r}\right)  e^{\Delta U(\mathbf{r},\mu_{c},\beta_{c}\mid n)}}{\int
d\mathbf{r}^{\prime}n^{(0)}\left(  \mathbf{r}^{\prime}\right)  e^{\Delta
U(\mathbf{r}^{\prime},\mu_{c},\beta_{c}\mid n)}}, \label{b.10}%
\end{equation}%
\begin{equation}
\Delta U(\mathbf{r},\mu_{c},\beta_{c}\mid n)=\int d\mathbf{r}^{\prime}\left(
c(\left\vert \mathbf{r-r}^{\prime}\right\vert ,\mu_{c},\beta_{c})n\left(
\mathbf{r}^{\prime},\mu_{c},\beta_{c}\right)  -c^{(0)}(\left\vert
\mathbf{r-r}^{\prime}\right\vert ,\mu_{c},\beta_{c})n^{(0)}\left(
\mathbf{r}^{\prime},\mu_{c},\beta_{c}\right)  \right)  . \label{b.11}%
\end{equation}
where $c_{T}(r,\mu_{c},\beta_{c})\rightarrow c^{(0)}(r,\mu_{c},\beta_{c})$
corresponding to the non-interacting limit. Clearly, $n\left(  \mathbf{r}%
,\mu_{c},\beta_{c}\right)  \rightarrow n^{(0)}\left(  \mathbf{r},\mu_{c}%
,\beta_{c}\right)  $ in the absence of Coulomb interactions. Although it is
not needed for calculation of (\ref{b.10}), the effective trap potential is
determined from%
\begin{equation}
\beta_{c}\left(  \mu_{ec}-v_{c}(\mathbf{r})\right)  ^{(0)}=\ln\left(
n^{(0)}\left(  \mathbf{r}\right)  \lambda_{c}^{3}\right)  +\int d\mathbf{r}%
^{\prime}c^{(0)}(\left\vert \mathbf{r-r}^{\prime}\right\vert ,\mu_{c}%
,\beta_{c})n^{(0)}\left(  \mathbf{r}^{\prime},\mu_{c},\beta_{c}\right)  .
\label{b.11a}%
\end{equation}
This is used in the calculations for Figure \ref{Fig.3}b. \ 

It is instructive to look at the limit of zero temperature. A Sommerfeld
expansion of the local density (\ref{b.9}) gives
\begin{equation}
n^{\ast(0)}\left(  \mathbf{r}^{\ast},t=0,r_{s}\right)  =\left\{
\begin{array}
[c]{c}%
0.034r_{s}^{3/2}\left(  \frac{2\nu_{0}}{\Gamma}-r^{\ast2}\right)
^{3/2},\hspace{0.2in}r^{\ast}<\sqrt{\frac{2\nu_{0}}{\Gamma}}\\
0,\hspace{0.2in}r^{\ast}\geq\sqrt{\frac{2\nu_{0}}{\Gamma}}%
\end{array}
\right.  , \label{b.12}%
\end{equation}
where $t\nu_{0}$ is determined from normalization
\begin{equation}
\nu_{0}=0.783\overline{N}^{1/3}\frac{r_{s}^{1/2}}{t},\hspace{0.2in}\frac
{2\nu_{0}}{\Gamma}=2.88\frac{\overline{N}^{1/3}}{r_{s}^{1/2}}. \label{b.12a}%
\end{equation}
The density is \ concave from the origin until $r^{\ast}=\sqrt{2.88\frac
{\overline{N}^{1/3}}{r_{s}^{1/2}}}$, beyond which it vanishes. This vanishing
of the density implies that the associated effective classical confining
potential develops a hard wall. For the case of Figure \ref{Fig.4},
$\overline{N}=100,r_{s}=18.4$, this gives $r^{\ast}\simeq1.77$. The shell
structure of Figures \ref{Fig.2} and \ref{Fig.4} are finite temperature
precursors of this limit.

With $n^{\ast(0)}\left(  \mathbf{r}^{\ast},t=0,r_{s}\right)  $ known, the
effective confining potential can be determined from (\ref{b.11}), where the
exact Fourier transform of the ideal gas direct correlation function has the
simple form \cite{Amovilli07}%
\begin{equation}
\widetilde{c}^{(0)}(k^{\ast},t=0,r_{s})=r_{0}^{3}\left(  1-\frac{1}{\frac
{3}{4k_{F}^{\ast}}k^{\ast}-\frac{1}{16k_{F}^{\ast}}k^{\ast3}}\right)  .
\label{b.12b}%
\end{equation}
Here $k_{F}^{\ast}=k_{F}r_{0}=(9\pi/4)^{1/3}$ and $k_{F}=(3\pi^{2}n)^{1/3}$ is
the Fermi wavelength.

\subsection{Weak Coulomb limit}

\label{apB2}The non-interacting limit of the previous subsection has only
exchange correlations among the particles to provide quantum effects on the
effective trap. A better limit, incorporating some mean field Coulomb
interactions as well is given by the weak Coulomb coupling approximation in
density functional theory (Hartree plus exchange). Within the same
Thomas-Fermi approximation as (\ref{b.9}) this is%
\begin{align}
n_{T}\left(  \mathbf{r},\mu_{c},\beta_{c}\right)   &  \rightarrow n^{\left(
w\right)  }(\mathbf{r},\mu_{c},\beta_{c})\equiv\frac{2}{h^{3}}\int
d\mathbf{p}\left(  e^{-\nu_{0}}e^{\left(  \beta\left(  \frac{p^{2}}%
{2m}+v(\mathbf{r})\right)  +\beta v^{\left(  w\right)  }(\mathbf{r})\right)
}+1\right)  ^{-1}\nonumber\\
&  =\lambda^{-3}\frac{4}{\sqrt{\pi}}I_{\frac{1}{2}}\left(  \left(  \nu
_{0}-\beta v(\mathbf{r})-\beta v^{\left(  w\right)  }(\mathbf{r})\right)
\right)  . \label{b.13}%
\end{align}
The potential $v^{\left(  w\right)  }(\mathbf{r})$\ representing the effects
of Coulomb interactions among the particles is given by%
\begin{equation}
v^{\left(  w\right)  }(\mathbf{r})=q^{2}\int d\mathbf{r}^{\prime}%
\frac{n^{\left(  w\right)  }\left(  \mathbf{r}^{\prime}\right)  }{\left\vert
\mathbf{r}-\mathbf{r}^{\prime}\right\vert }+v_{x}(n^{\left(  w\right)
}\left(  \mathbf{r}\right)  ). \label{b.14}%
\end{equation}
The first term is the mean-field Coulomb contribution (Hartree), while the
second term $v_{x}(n\left(  \mathbf{r}\right)  )$ is the local density
approximation for exchange (density derivative of the exchange free energy
\cite{Perrot79})%
\begin{equation}
v_{x}(n\left(  \mathbf{r}\right)  )=-\frac{e^{2}}{\sqrt{\pi}\lambda}%
I_{-\frac{1}{2}}\left(  \nu_{0}\left(  \mathbf{r}\right)  \right)  .
\label{b.15}%
\end{equation}
The density dependence of $v_{x}\left(  n\right)  $ is determined by inverting
the ideal gas relationship%
\begin{equation}
n(\mathbf{r})=\lambda^{-3}\frac{4}{\sqrt{\pi}}I_{\frac{1}{2}}\left(  \nu
_{0}(\mathbf{r})\right)  . \label{b.16}%
\end{equation}

It remains to determine the corresponding approximation to the classical
direct correlation function, $c_{T}\rightarrow c^{\left(  w\right)  }$. Since
(\ref{b.14}) results from an expansion of the Kohn-Sham potential to leading
order in the Coulomb coupling constant $\Gamma$, the function $c^{\left(
w\right)  }$ is the corresponding weak coupling (small $\Gamma$) limit of $c$%
\begin{equation}
c^{\left(  w\right)  }(\left\vert \mathbf{r-r}^{\prime}\right\vert ,\mu
_{c},\beta_{c})=c^{(0)}(\left\vert \mathbf{r-r}^{\prime}\right\vert ,\mu
_{c},\beta_{c})+\Gamma c^{(1)}(\left\vert \mathbf{r-r}^{\prime}\right\vert
,\mu_{c},\beta_{c}), \label{b.17}%
\end{equation}
and accordingly $\Delta U(\mathbf{r},\mu_{c},\beta_{c}\mid n)$ in (\ref{b.11})
becomes%
\begin{align}
\Delta U(\mathbf{r},\mu_{c},\beta_{c}  &  \mid n)\rightarrow\int
d\mathbf{r}^{\prime}\left(  \left(  c(\left\vert \mathbf{r-r}^{\prime
}\right\vert ,\mu_{c},\beta_{c})n\left(  \mathbf{r}^{\prime},\mu_{c},\beta
_{c}\right)  \right)  \right. \nonumber\\
&  \left.  -\left(  c^{(0)}(\left\vert \mathbf{r-r}^{\prime}\right\vert
,\mu_{c},\beta_{c})+\Gamma c^{(1)}(\left\vert \mathbf{r-r}^{\prime}\right\vert
,\mu_{c},\beta_{c})\right)  n^{\left(  w\right)  }\left(  \mathbf{r}^{\prime
},\mu_{c},\beta_{c}\right)  \right)  \label{b.18}%
\end{align}
The analytic calculation of $c^{(1)}$ from expansion in $\Gamma$ does not lead
to a simple, practical result. Instead, it can be calculated numerically from
the HNC equations using a small value for $\Gamma$ and writing
\begin{equation}
c^{(1)}(r,\mu_{c},\beta_{c})=\lim\frac{1}{\Gamma}\left(  c(r,\mu_{c},\beta
_{c})-c^{(0)}(r,\mu_{c},\beta_{c})\right)  . \label{b.19}%
\end{equation}
In terms of the variables $t,r_{s}$ the notion of small $\Gamma$ is ambiguous%

\begin{equation}
\Gamma=\frac{\beta q^{2}}{r_{0}}=\frac{r_{s}}{t}\frac{2}{\left(  \frac{9}%
{4}\pi\right)  ^{2/3}}, \label{b.20}%
\end{equation}
However, since the non-interacting case depends only on $t$ the charge
coupling can be considered the effect which introduces the $r_{s}$ dependence.
Hence $\Gamma$ should be made small by choosing the appropriate values for
$r_{s}<<1$ . Then $c^{(1)}$ will be a function of $t$ alone.

In summary, with the limit density $n^{\left(  w\right)  }\left(
\mathbf{r},\mu_{c},\beta_{c}\right)  $ and $\Delta U(\mathbf{r},\mu_{c}%
,\beta_{c}\mid n)$ given by (\ref{b.18}) the dimensionless forms (\ref{4.1})
\ and (\ref{4.2}) of the text are obtained. If desired, the effective trap can
be calculated from (\ref{b.4}) which becomes
\begin{align}
\left(  \beta\mu_{ec}-\beta v_{c}(\mathbf{r},\mu_{c},\beta_{c})\right)
^{\left(  w\right)  }  &  =\ln\left(  n^{\left(  w\right)  }\left(
\mathbf{r},\mu_{c},\beta_{c}\right)  \lambda_{c}^{3}\right) \nonumber\\
&  -\int d\mathbf{r}^{\prime}\left(  c^{(0)}(\left\vert \mathbf{r-r}^{\prime
}\right\vert ,\mu_{c},\beta_{c})+\Gamma c^{(1)}(\left\vert \mathbf{r-r}%
^{\prime}\right\vert ,\mu_{c},\beta_{c})\right)  n^{\left(  w\right)  }\left(
\mathbf{r}^{\prime},\mu_{c},\beta_{c}\right)  . \label{b.21}%
\end{align}

\section{Validity of Thomas-Fermi forms}

\label{apC}

\begin{figure}[ptb]
\includegraphics[width=\columnwidth]{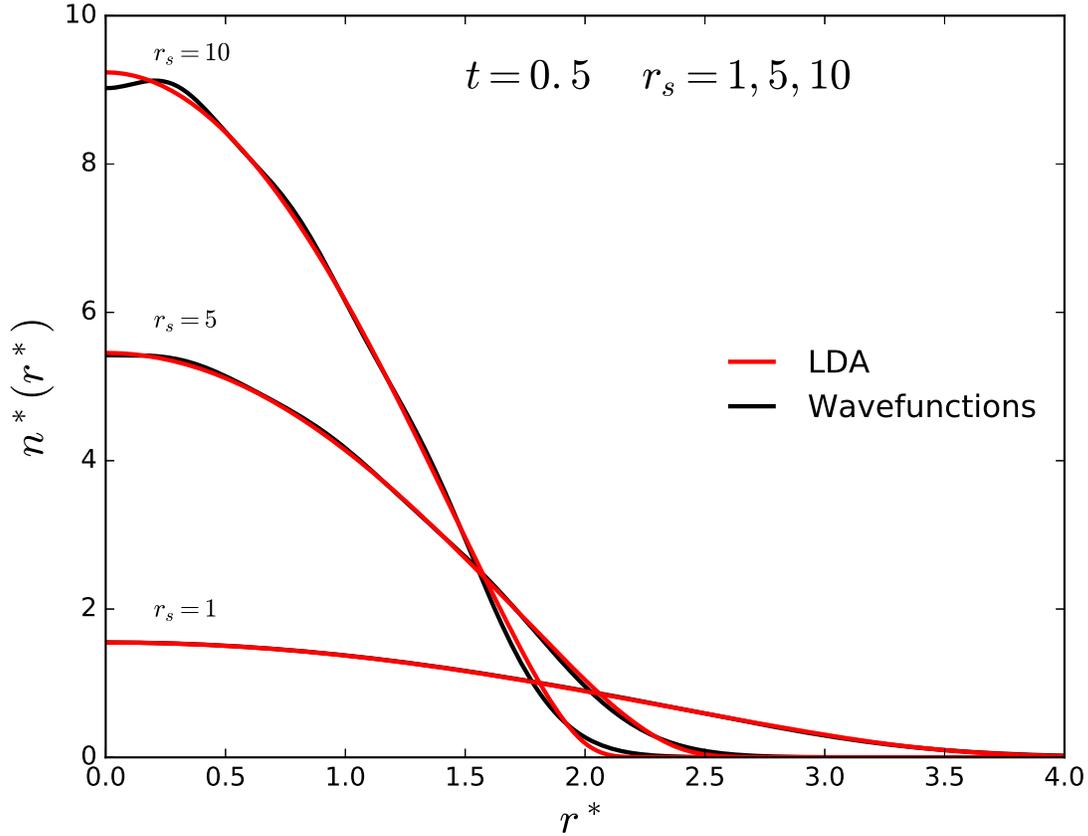}\caption{Comparison of density profiles calculated with Harmonic oscillator wavefunctions (black lines) and with the Thomas-Fermi approximation (red lines) for $t=0.5$ and $r_s=1, 5, 10$.}%
\label{Fig.8}%
\end{figure}

Consider again (\ref{b.8}) for the non-interacting density
\begin{equation}
n^{\ast(0)}\left(  r,t,r_{s}\right)  =r_{0}^{\ast3}\sum_{\alpha}\left\vert
\psi_{\alpha}\left(  \mathbf{r}\right)  \right\vert ^{2}\left(  e^{\beta
\left(  \epsilon_{\alpha}-\mu_{c}\right)  }+1\right)  ^{-1}. \label{c.1}%
\end{equation}
and its Thomas-Fermi (local density) approximation (\ref{b.9})
\begin{equation}
n^{\ast(0)}\left(  r,t,r_{s}\right)  \simeq\left(  \frac{r_{0}^{\ast}}%
{\lambda}\right)  ^{3}\frac{4}{\sqrt{\pi}}I_{\frac{1}{2}}\left(  \left(
\beta\mu_{e}-\frac{1}{2}\Gamma r^{\ast2}\right)  \right)  . \label{c.2}%
\end{equation}
Both are normalized to $\overline{N}=100$. Figure \ref{Fig.8} shows their
comparison at $t=0.5$ for $r_{s}=1,5,10$. The agreement is quite good even for
this low temperatures. Normally one would expect the Thomas-Fermi form to be
applicable only at temperatures well above the Fermi temperature and for
smooth densities. Evidently the large particle number considered here has
extended its validity to lower temperatures.

\bigskip
\end{document}